\documentclass[varenna,seceqno]{cimento}

\usepackage{amsmath}

\title{Gravito-Magnetism in one-body and two-body systems: Theory
and Experiments.}

\author{R. F. O'Connell}

\institute{Department of Physics and Astronomy, Louisiana State
University, \\ Baton Rouge, Louisiana  70803-4001, USA}

\begin{document}

\maketitle

\begin{abstract}
We survey theoretical and experimental/observational results on
general-relativistic spin (rotation) effects in binary systems.  A
detailed discussion is given of the two-body Kepler problem and
its first post-Newtonian generalization, including spin effects. 
Spin effects result from gravitational spin-orbit and spin-spin
interactions (analogous to the corresponding case in quantum
electrodynamics) and these effects are shown to manifest
themselves in two ways: (a) precession of the spinning bodies per
se and (b) precession of the orbit (which is further broke down
into precessions of the argument of the periastron, the longitude
of the ascending node and the inclination of the orbit).  We also
note the ambiguity that arises from use of the terminology
frame-dragging, de Sitter precession and Lense-Thirring
precession, in contrast to the unambiguous reference to
spin-orbit and spin-spin precessions.
Turning to one-body experiments, we discuss the recent results of
the GP-B experiment, the Ciufolini-Pavlis Lageos experiment and
lunar-laser ranging measurements (which actually involve three
bodies).  Two-body systems inevitably involve astronomical
observations and we survey results obtained from the first binary
pulsar system, a more recently discovered binary system and,
finally, the highly significant discovery of a double-pulsar
binary system.
\end{abstract}

\section{Introduction }

This manuscript is, in essence, a progress report on
general-relativistic spin effects in binary systems, which I
lectured on in Varenna summer schools in 1974 \cite{oconnell74} and
1975 \cite{barker76}.  Here, we use "spin" in the generic sense of
meaning "internal spin" in the case of an elementary particle and
"rotation" in the case of a macroscopic body.  My 1974 lecture
concentrated on the one-body problem
\cite{barker70} and analysis of the Schiff gyroscope experiment
\cite{schiff}, now referred to as the Gravity Probe B (GP-B)
experiment
\cite{everitt}, which was launched on April 20, 2004 \cite{brum}. 
In addition, I discussed the possibility of C, P, and T effects
occuring in the gravitational interaction.  Not long after the 1974
summer school, a momentous event occured: the discovery of a pulsar
in a binary system by Hulse and Taylor \cite{hulse} which bought
into play a new array of possibilities for observing general
relativistic effects.  In particular, this motivated Barker and me
to calculate two-body spin precession effects
\cite{barker75, barker1975}, more details which were reported on
later at the 1975 Varenna summer school.  Thus, other than
summarizing the key results from these lectures, I will
concentrate here on discussing progress since then, in theory,
experiment and observation.  Section 2 is devoted to a general
discussion of the Kepler problem and its first post-Newtonian
generalization (order $c^{-2}$) including spin effects. In section
3, we discuss a one-body system and related experiments and
observations, not only the GP-B experiment but also the more recent
results of Ciufolini and Pavlis \cite{ciu} and the very early
theoretical work of de Sitter \cite{sitter76, sitter72} and
Lense-Thirring \cite{lense1, lense2} as well as the related work of
Williams et al.
\cite{will3} and Murphy et al. \cite{murphy07}.  Section 4
concentrates on 2-body systems and we discuss, in particular, the
observations of Stairs et al. \cite{stairs} and the momentous
discovery of a double binary system \cite{lyne}.

\section{The Kepler Problem and its Post-Newtonian Generalization}

For a binary system, the Newtonian gravitational potential gives
rise, for energy $E<0$, to elliptic motion.  We are interested in
post-Newtonian generalizations (order $c^{-2}$) arising from both
spin and general relativistic effects.  First, we present our
notation \cite{barker76}.  Let $m_{1}, \mathbf{r}_{1},
\mathbf{v}_{1}, \mathbf{P}_{1},
\mathbf{S}^{(1)}, \mathbf{n}^{(1)}, I^{(1)}$ and
$\boldsymbol{\omega}^{(1)}$ denote the mass, position, velocity,
momentum, spin, unit vector in the spin direction, moment of
inertia and angular velocity, respectively, of body 1.  The same
symbols, with $1\rightarrow{2}$, denote the corresponding
quantities for body 2.  In the center-of-mass system we have
$\mathbf{r}=\mathbf{r}_{1}-\mathbf{r}_{2}$ and
$\mathbf{P}=\mathbf{P}_{1}= -\mathbf{P}_{2}=\mu\mathbf{v}$ where
$\mathbf{v}=\mathbf{v}_{1}-\mathbf{v}_{2}$.  The reduced mass and
the total mass are given by

\begin{equation}
\mu\equiv\frac{m_{1}m_{2}}{m_{1}+m_{2}},~~~~~~~M\equiv{m}_{1}+m_{2}.
\label{gm21}
\end{equation}  Also, the orbital angular momentum is

\begin{equation}
\mathbf{L}=\mathbf{r}\times\mathbf{P}\equiv{L}\mathbf{n},
\label{gm22}
\end{equation} where $\mathbf{n}$ is a unit vector in the
$\mathbf{L}$-direction. In additon,

\begin{equation}
\frac{L/\mu}{a^{2}(1-e^{2})^{\frac{1}{2}}}=\left(\frac{GM}{a^{3}}\right)
^{\frac{1}{2}}=\frac{2\pi}{T}=\bar{\omega}, \label{gm23}
\end{equation} where $T$ is the orbital period, $e$ is the
eccentricity,
$a$ is the semi-major axis, and $\bar{\omega}$ is the average
orbital angular velocity.

For a Newtonian elliptic orbit for two spherically symmetric
bodies, the energy $E$, the orbital angular momentum $\mathbf{L}$,
and the Runge-Lenz vector $\mathbf{A}$ are constants of the motion,
which can be written as

\begin{equation} E/\mu=\frac{1}{2}v^{2}-GM/r, \label{gm24}
\end{equation}

\begin{equation}
\mathbf{L}/\mu=\mathbf{r}\times\mathbf{v}, \label{gm25}
\end{equation} and

\begin{equation}
\mathbf{A}/\mu=\mathbf{v}\times(\mathbf{r}\times\mathbf{v})-GM\mathbf{r}/
r=\frac{1}{\mu}(\mathbf{v}\times\mathbf{L})-GM\mathbf{r}/r,
\label{gm26}
\end{equation}  where the direction of $\mathbf{A}$ is along the
major axis from the focus to the perihelion and is thus
perpindicular to $\mathbf{L}$.  However, when spin and
post-Newtonian effects are taken into account (order
$c^{-2}$ beyond Newtonian theory) we found that the secular results
for the precession of the orbit are

\begin{equation}
\dot{E}_{av}=0, \label{gm27}
\end{equation}

\begin{equation}
\dot{\mathbf{L}}_{av}=\boldsymbol{\Omega}^{\ast}\times\mathbf{L},
\label{gm28}
\end{equation}

\begin{equation}
\dot{\mathbf{A}}_{av}=\boldsymbol{\Omega}^{\ast}\times\mathbf{A},
\label{gm29}
\end{equation} where

\begin{equation}
\boldsymbol{\Omega}^{*}=\boldsymbol{\Omega}^{*(E)}+\boldsymbol{\Omega}^{*(1)}+
\boldsymbol{\Omega}^{*(2)}+\boldsymbol{\Omega}^{*(1,2)},
\label{gm210}
\end{equation} consists of contributions from a spin-independent
term $\boldsymbol{\Omega}*^{(E)}$ (Einstein's precession of the
perihelion), in addition to $\mathbf{S}^{(1)},
\mathbf{S}^{(2)}~and~\mathbf{S}^{(1)}
\mathbf{S}^{(2)}$ terms, respectively, which are all given
explicitly in
\cite{barker75}.  We note that, on the average,
$\mathbf{L}$ and
$\mathbf{A}$ precess at the same rates, so that the Kepler ellipse
as a whole precesses with the angular velocity
$\boldsymbol{\Omega}^{*}$.  It was also found convenient (in order
to adhere to astronomical and space physics practise) to write
$\bf{\Omega}^{\ast}$ in the form

\begin{equation}
\boldsymbol{\Omega}^{\ast}=\frac{d\Omega}{dt}\mathbf{n}_{0}+\frac{d\omega}{dt}
\mathbf{n}+\frac{di}{dt}\frac{\mathbf{n}_{0}\times\mathbf{n}}{|\mathbf{n}_{0}
\times\mathbf{n}|}, \label{gm211}
\end{equation} where $\Omega, \omega~and~i$ denote the longitude
of the ascending node, the argument of the periastron and the
inclination of the orbit, respectively, in the reference system of
the plane of the sky (the tangent plane to the celestial sphere at
the center of mass of the binary system) \cite{barker76,
barker70}.  In addition, $\mathbf{n}_{0}$ is a unit vector normal
to the plane of the sky directed from the center of mass of the
binary system towards the Earth.  The angle between
$\mathbf{n}_{0}$ and $\mathbf{n}$ is the inclination $i$.  In the
absence of spin, only the periastion precession is present.

In addition, each spin precesses at rates $\Omega^{(1)}$ and
$\Omega^{(2)}$, which when averaged over an orbital period we
write as

\begin{equation}
\dot{\mathbf{S}}^{(1)}_{av}=\boldsymbol{\Omega}^{(1)}_{av}\times\mathbf{S}^{(1)},
\label{gm212}
\end{equation} and similarly for $1\rightarrow{2}$.  However, as we
recently stressed "gravitational effects due to rotation (spin)
are best described, using the language of QED, as spin-orbit and
spin-spin effects since they also denote the interactions by which
such effects are measured; in fact these are the only such spin
contributions to the basic Hamiltonian describing the gravitational
two-body system with arbitrary masses, spins and quadruple
moments.  They manifest themselves in just two ways, spin and
orbital precessions, and whereas these can be measured in a variety
of ways (for example, as discussed above, orbital precession can be
subdivided into periastron, nodal and inclination precessions)
such different measurements are simply 'variations on the theme'"
\cite{oconnell05}.  If fact, in the basic Hamiltonian for the
problem
\cite{barker75}, the spin-orbit terms are of the form
$\mathbf{S}^{(1)}\cdot\mathbf{L}$ and
$\mathbf{S}^{(2)}\cdot\mathbf{L}$ whereas the spin-spin terms are
of the form $
\{[3(\mathbf{S}^{(1)}\cdot\mathbf{r})(\mathbf{S}^{(2)}\cdot\mathbf{r})/r^{2}
-\mathbf{S}^{(1)}\cdot\mathbf{S}^{(2)}]\}$, which is completely
analogous to what occurs in QED, as we pointed out in
\cite{oconnell74}.  Thus, we were able to conclude
\cite{oconnell74} that all such effects in QED have their analogy
in general relativity and, except for (important) numerical
factors, the latter results may be obtained from the former by
simply letting
$e^{2}\rightarrow{G}m_{1}m_{2}$. Hence it is convenient to write
(with a similar equation for 1 $\rightarrow$ 2)

\begin{eqnarray}
\boldsymbol{\Omega}^{(1)}_{av} &=&
\boldsymbol{\Omega}^{(1)}_{so}+\boldsymbol{\Omega}^{(1)}_{ss}
\nonumber \\ &\equiv&
\boldsymbol{\Omega}^{(1)}_{ds}+\boldsymbol{\Omega}^{(1)}_{LT},\label{213}
\end{eqnarray} where the second equality is often written to make
contact with common usage, $ds$ denoting de Sitter
\cite{sitter76,sitter72} and $LT$ denoting Lense-Thirring
\cite{lense1,lense2}.  The former is also referred to as the
geodetic effect and the latter as a "frame-dragging" effect. 
However, as we pointed out in
\cite{oconnell05}, the de Sitter precession can also be regarded as
a frame-dragging effect, which is even more obvious in the case of
two-body systems (but which is also readily seen in the case of the
GP-B experiment if one considers an observer in the frame of the
gyroscope). Further confusion associated with the terminology
"Lense-Thirring effect" arises from the fact that it is also used
to discuss earth spin effects on orbital motion, as we discuss in
more length below (subsect. 3.3).  Thus, to avoid confusion, we
feel that it is best to just use the terminolgy given in the
first equality of (2.13).  We also note that
$\boldsymbol{\Omega}^{(1)}_{so}$ and
$\boldsymbol{\Omega}^{(2)}_{so}$ (explicit results for which are
given in
\cite{barker75}) do not depend on the spins and always point along
$\vec{L}$ whereas
$\boldsymbol{\Omega}^{(1)}_{ss}$ depends on the spin of body 2 and
points along
$\left\{\left[\mathbf{n}^{(2)}-3\left(\mathbf{n}\cdot
\mathbf{n}^{(2)}\right)\mathbf{n}\right]\right\}$.

To conclude this section, we make some general remarks.  First, we
note that since gravitational radiation only arises at order
$c^{-5}$, and defining the total angular momentum

\begin{equation}
\mathbf{J}\equiv\mathbf{L}+\mathbf{S}^{(1)}+\mathbf{S}^{(2)},
\label{gm214}
\end{equation} we obtained \cite{barker75}

\begin{equation}
\frac{d\mathbf{J}}{dt}=0. \label{gm215}
\end{equation}  again underlining the inter-relationship between
the spin and orbital precessions.  Secondly, we note that the
largest precession is always the periastion precession and it is
the only one that remains in the absence of spin.  For a particular
system, the magnitude of the periastion precession and the
spin-orbit precessions depends significantly on the value of the
product of the orbital angular velocity $\bar{\omega}$ and the
gravitational coupling constant

\begin{equation}
\alpha_{g}=(GM/c^{2}a). \label{gm216}
\end{equation}  where $a$ is the semi-major axis of the Keplerian
ellipse.  Thus, for the GP-B satellite orbiting the earth,
$\bar{\omega}=1.074\times 10^{-3}s^{-1}$ and
$\alpha_{g}=6.958\times 10^{-10}$, so that
$\alpha_{g}\bar{\omega}=7.473\times 10^{-13}s^{-1}=
\left(1.35\times 10^{-4}\right)^{0}/yr$. On the other hand, for the
double binary system,
$\bar{\omega}=7.112\times 10^{-4}s^{-1}$ and
$\alpha_{g}=4.348\times 10^{-6}$ so that
$\alpha_{g}\bar{\omega}=3.092\times 10^{-9} s^{-1}=5^{o}.591/yr$. 
As we will see below, this difference, amounting to 4 orders of
magnitude, is reflected in the significantly larger precessions
(both orbital and spin) obtained for binary star systems compared
to earth related systems.  For example, this order-of-magnitude
calculation is borne out by an exact calculation of the spin-orbit
precession, which gives (in degrees per year)
$1.84\times 10^{-3}$ for the GP-B gyroscope and 4.8 and 5.1 for
pulsars A and B, respectively, which constitute the double binary
system, to be discussed below.  Thirdly, the inter-relationship
between the spin and orbital precessions, as expressed in
(\ref{gm215}), may be seen explicitly by writing our exact basic
results
\cite{barker75} in the following succient form (recalling again
that the precession rates are averaged over an orbital period)

\begin{eqnarray}
\frac{d\mathbf{S}^{(1)}}{dt} &=&
\left(\boldsymbol{\Omega}^{(1)}_{so}+\boldsymbol{\Omega}^{(1)}_{ss}\right)\times
\mathbf{S}^{(1)} \nonumber \\ &=&
\frac{3}{2(1-e^{2})}~\alpha_{g}\bar{\omega}~\frac{m_{2}+\mu
/3}{M}~~~\left(\mathbf{n}\times
\mathbf{S}^{(1)}\right) \nonumber \\
&&+\frac{1}{2(1-e^{2})}~\alpha_{g}~\frac{S^{(2)}}{Ma^{2}}~\left[\left(
\mathbf{n}^{(2)}\times\mathbf{S}^{(1)}\right)-3\left(\mathbf{n}\cdot\mathbf{n}^{(2)}\right)
\left(\mathbf{n}\times\mathbf{S}^{(1)}\right)\right],
\label{gm217}
\end{eqnarray}

\begin{eqnarray}
\frac{d\mathbf{S}^{(2)}}{dt} &=&
\left(\boldsymbol{\Omega}^{(2)}_{so}+\boldsymbol{\Omega}^{(2)}_{ss}\right)\times
\mathbf{S}^{(2)} \nonumber \\ &=&
\frac{3}{2(1-e^{2})}~\alpha_{g}\bar{\omega}~\frac{m_{1}+\mu
/3}{M}~~~\left(\mathbf{n}\times
\mathbf{S}^{(2)}\right) \nonumber \\
&&+\frac{1}{2(1-e^{2})}~\alpha_{g}~\frac{S^{(1)}}{Ma^{2}}~\left[
\left(\mathbf{n}^{(1)}\times\mathbf{S}^{(2)}\right)-3\left(\mathbf{n}\cdot\mathbf{n}^{(1)}\right)
\left(\mathbf{n}\times\mathbf{S}^{(2)}\right)\right],
\label{gm218}
\end{eqnarray} and

\begin{eqnarray}
\frac{d\mathbf{L}}{dt} &=&
\left(\boldsymbol{\Omega}^{*(E)}+\boldsymbol{\Omega}^{*(1)}+\boldsymbol{\Omega}^{*(2)}+
\boldsymbol{\Omega}^{(1,2)}
\right)\times\mathbf{L} \nonumber \\ &=&
-\left\{\left(\boldsymbol{\Omega}^{(1)}_{so}\times\mathbf{S}^{(1)}\right)+\left(\boldsymbol{\Omega}^{(2)}_{so}\times
\mathbf{S}^{(2)}_{so}\right)+\left[\left(\boldsymbol{\Omega}^{(1)}_{ss}\times\mathbf{S}^{(1)}\right)+\left(
\boldsymbol{\Omega}^{(2)}_{ss}\times\mathbf{S}^{(2)}\right)\right]\right\}
\nonumber \\
&=&-\left(\frac{d\mathbf{S}^{(1)}}{dt}+\frac{d\mathbf{S}^{(2)}}{dt}\right),
\label{gm219}
\end{eqnarray} which is consistent with (\ref{gm214}) and
(\ref{gm215}).  This is another manifestation of the fact that
there are simply two basic gravito-magnetic interactions in general
relativity, spin-orbit and spin-spin.  We note that
$\boldsymbol{\Omega}^{*(E)}$ does not contribute to
$\frac{d\mathbf{L}}{dt}$ whereas for $\frac{d\mathbf{A}}{dt}$ it
is the dominant Einstein-Robertson periastion precession term. 
It is also immediately clear from these results that the
various precession rates are proportional to $\alpha_{g}$ times the
average orbital rate, consistent with the fact that we are
working to order $c^{-2}$ (first post-Newtonian
approximation). After this general discourse, we now turn to a
discussion of various experiments to detect  spin-orbit and
spin-spin effects in gravitational theory.

\section{One-Body System}

\subsection{The GP-B experiment}

As already mentioned, it was launched on April 20, 2004.  Data
collection covered the period Aug. 2004-Sept. 2005, when the He in
the dewar had completely leaked out but, due to unexpected
"--measurement uncertainty--," the announcement of final results
has been delayed until December 2007.

Here $m_{2}$ (earth mass) $>>m_{1}$ (gyroscope mass) and
$\mathbf{S}^{(2)}$ is along the earth's rotation axis.  Since
$\boldsymbol{\Omega}^{(1)}$ is always along
$\mathbf{L}$, it turns out that a polar orbit results in
$\boldsymbol{\Omega}^{(1)}_{so}$ and
$\boldsymbol{\Omega}^{(1)}_{ss}$ being at right angles resulting,
in principle, in a clear separation of the two effects.  However,
due to a variety of other effects
\cite{barker84,oconnell85}, the analysis turns out to be much more
complicated.  The calculated precessional rates for the GP-B
experiment, in units of arc-seconds per year ($2.78\times 10^{-4}$
degrees per year) are 6.61 for $\Omega^{(1)}_{so}$ and
$4.1\times 10^{-2}$ for $\Omega^{(2)}_{ss}$ and the hope is that
the accuracy achieved is better than $5\times 10^{-4}$ ($1.4\times
10^{-7}$ degrees) per year
\cite{kahn07}.  However, whether this is achievable in view of
the problems encountered \cite{brum,kahn07,stanford06} remains
to be seen.  Among the concerns is the polhode motion arising from
the fact that the gyroscopes are not perfect sphere.  Thus, not only
can a quadrupole moment arise from the manufacturing process but,
in addition, even an ideal perfectly spherical gyro acquires a
quadrupole moment due to its rotation \cite{barker84}.  Using
the quoted values of the gyro diameter and average spin rate
\cite{kahn07} we obtain a value for ($\Delta$I/I), arising from
this effect, of $7.1\times 10^{-6}$, which is comparable to that
arising from the manufacturing process.  An additional problem as
far as the experiment is concerned is that the two different
quadrupole moments are located at an unknown angle from each other,
aggravated by the fact that "--there was no way to know about which
axis each gyro (of the four total) was initially spun up--"
\cite{stanford06}.  Moreover, the "--wide variation in spin-down
rates--" will be reflected by a time variation in the value of
($\Delta$I/I) associated with rotation (since $\Delta$I is
proportional to the square of the gyro's angular velocity), which
leads to a time variation in the polhode periods.  In addition, we
note that the two different quadrupole moments also affect the
orbit equations resulting in a concomitant contribution to the
spin precession rate \cite{barker84}.  The inter-connection of
these various parameters presents a challenge to the
interpretation of the data, particularly for the determination of
$\Omega_{ss}$, which is 161 times smaller than
$\Omega_{ss}$.  The data analysis is further aggravated
\cite{kahn07,stanford06} by the fact that unexpected
classical torques are produced on the gyros, together
with damping of their polhode motion, which are caused
by interactions between the gyro rotors and their housings due to
electrostatic patches (potential differences).

Since nearly five decades has passed since the GP-B experiment was
conceived and designed, there has been much progress in achieving
the same results by a variety of other methods which we will
discuss below.  In particular, there are now three independent
determinations of $\Omega_{so}$ consistent with the predicted
results arising from Einstein's theory and, in fact, one of them
pertains to a 2-body system for which additional interesting
features arise.  Also, whereas it is reported that GP-B has
confirmed the predicted spin-orbit result to better than 1\%
\cite{kahn07}, Murphy et al. 
\cite{murphy07} are claiming 10 times greater accuracy based on
lunar lasing. Thus, more attention will be focussed on the main
"selling-point" of the experiment, that is the determination of
$\Omega_{ss}$.  But, whether this is achievable is very
problematic in view of the polhode
 problems already mentioned and other various classical torques on
the gyroscopes \cite{kahn07}.
  Of course, there have been many significant "spin-offs"
connected with the experiment.  But, whether or not this will be
regarded as justification for the time spent and the cost (over
\$760 million) and in view of the fact that it required the U.S.
Congress to keep it alive after NASA "--has tried to cancel the
mission at least three times--"
\cite{reichhardt03,wynne06} will surely give rise to an analysis
of the sociology and politics of science research and funding,
along the lines that Collins presented for the search for
gravitational waves \cite{collins04}.

\subsection{Determination of $\Omega_{so}$ from
lunar-laser-ranging measurements}

As already alluded to above and discussed in detail in
\cite{oconnell05}, the de Sitter precession $\Omega_{so}
(\Omega_{ds})$ is also a frame-dragging effect and "--provides an
accurate benchmark measurement of spin-orbit effects which GPB
needs to emulate--." \cite{oconnell05}.  In essence, starting with
the work of de Sitter and investigated in detail by many groups
\cite{bertotti87,will3,murphy07}, it is clear that "--because of
its distance from the sun, the earth-moon system can be regarded as
a single body which is rotating in the gravitational field of the
sun.  In other words, the earth-moon system is essentially a
gyroscope in the field of the sun and its frame-dragging effect due
to interaction with the sun has been measured, using lunar laser
ranging--" \cite{oconnell05}.  First we note that the relevant
$\alpha_{g}$ is
$9.8724\times 10^{-9}$ and $\bar{\omega}=1.991\times
10^{-7}s^{-1}$.  The product
$\frac{3}{2} \alpha_{g}\bar{\omega}$ results in a de Sitter
\cite{sitter76,sitter72} spin-orbit contribution which is
reflected in a contribution to the lunar perigee
$\approx$ 19.2 m-sec/yr
\cite{bertotti87}.  Second, and more recently, Murphy et al.
\cite{murphy07} pointed out that the current accuracy of lunar
lasing ranging is such that it provides a test of spin-orbit
coupling to "--approximately 0.1\% accuracy - better than the
anticipated accuracy of the gravity - Probe B result--" and that
"--a new effort in LLR is poised to deliver order-of-magnitude
improvements in range precision".

\subsection{Ciufolini's Lageos experiment}

The fundamental idea for this work goes back to Ciufolini's work
in mid 1980, details of which may be found in \cite{ciufolini86}. 
In their recent paper
\cite{ciu,ciufolini04}, Ciufolini and Pavlis used in their
analysis two satellites, Lageos (launched in 1976) and Lageos2
(launched in 1992) which simply consist of a heavy sphere of
retro-reflectors which reflect short laser pulses sent from earth,
leading to a determination of earth-satellite distances with a
precision of a few
$mm$.  The focus is on the orbital  motion of the satellites and
specifically a determination of $d\Omega /dt$, the nodal
precession, due to the spin of the earth.  As already emphasized,
this is a spin-orbit effect, involving the spin of the earth and
the angular momentum of the orbit; the only difference with the
corresponding spin-orbit measurement in the GP-B experiment is
that the spin under discussion there is the spin of the gyroscope
(the spin of the earth coming into play only in the spin-spin
interaction involving the earth spin and the gyroscope spin).  Our
emphasis on precise terminology stems from the fact that there is
confusion associated with the use of the terminology
"Lense-Thirring effect"; some authors, such as \cite{ciu}, use it
when the earth's spin is playing a role whereas others, such as
\cite{everitt,kahn07} use it to denote the interaction between
the earth's spin and the spin of the gyroscope.  The emphasis in
\cite{ciu} on nodal motion is based on the fact that, in this case,
non-gravitational perturbations are easiest to handle.  Also, the
use of two satellites made it possible to eliminate the effect of
the earth's quadrupole moment.  The use of the GRACE earth gravity
model
\cite{ciufolini04}, resulting in very small uncertainties arising
from other harmonics, led to a value of
$48.2 masyr^{-1}$ for the nodal precession, accurate to a precision
of 10\%.  This is a very impressive result and the error estimate
is bound to become even smaller as better results emerge for the
various multipoles of the earth's field and for the expected
improvements in laser-ranging precision.  For an excellent review
of recent measurements of frame dragging using earth-orbiting
satellites, we refer to Ciufolini's Nature review
\cite{ciufolini07}, where he remarks that the laser-ranged Italian
satellite LARES should " - - in future provide an improved test of
the Earth's gravitomagnetism with accuracy of the order of
1$\%$".

\section{Two-Body Systems}

\subsection{The binary pulsar PSR 1913+16 \cite{hulse}}

It was the first discovery of a pulsar in a binary system and
stimulated Barker and me to extend our one-body analysis
\cite{barker70} to the two-body arena
\cite{barker75}, the results of which were immediately applied to
the Hulse-Taylor system \cite{barker1975}.

In the two-body system, new and unexpected results were obtained. 
The only previous two-body calculation was that of Robertson
\cite{robertson38} who neglected spin and considered only two test
particles for which he calculated the periastion precession, the
result for which was the same as the one-body calculation except
for the replacement $m_{2}\rightarrow m_{1}+m_{2}\equiv M$. 
However, when spin is taken into account, the situation is very
different.  What we found is that, for the spin precession of body
1, the replacement in the formula for $\Omega^{(1)}_{so}$ is
$m_{2}\rightarrow m_{2}+(\mu /3)$ and, in the case of $\Omega
^{(2)}_{so}$, the replacement is $m_{1}\rightarrow m_{1}+(\mu
/3)$, where we recall, from (\ref{gm21}), that $\mu$ is the reduced
mass.  Similar replacements occur for the orbital precessions
$\Omega^{*(1)}$ and $\Omega^{*(2)}$, associated with the
spin-orbit interaction contributions from the spins of bodies 1 and
2, respectively. This is shown explicitly in (\ref{gm217}) to
(\ref{gm219}). Here
$\alpha_{g}=2\times 10^{-6}$, compared to a value of $7\times
10^{-10}$ for the earth-gyro system and thus we expect
correspondingly large numbers for the various precession angles. 
In line with these expectations, the observations give
$\Omega^{*(E)}=4^{0}.2/yr$, in agreement with theory.  The
theoretical value for
$\Omega^{(1)}_{so}$ is $1^{0}.1/yr$ (where $m_{1}=1.42 m_{\odot}$
 refers to the spinning pulsar) but this quantity is difficult to
deduce from the data. Two obvious ways suggest themselves: (a) wait
for the pulsar beam (which is directed in a narrow cone toward the
earth) to go outside the line of sight, (which could take a long
time since the precessional period is about 327 years), or (b)
observe the changes over time in the shape of the pulses, which
have now been seen
\cite{kramer98,weisberg02} and which give qualitative agreement
with theory but due to uncertainties in the beam shape, a
quantitative comparison has not yet been possible.

\subsection{The binary pulsar PSR B1534+12 \cite{stairs}}

Here, based on the two-body spin precession formula given in
\cite{barker75} and
\cite{barker1975}, the theoretical precession rate is $0.51
^{0}/yr$.  Observations amounting to over 400 hours led to a
determination of the time evolution of (a) the angle between the
spin and the magnetic pole and (b) the minimum impact angle of the
magnetic pole on the line of sight, from which the authors deduced
direct evidence of geodetic (spin-orbit) precession.  This was
bolstered by an independent determination based on time variation
of the shape of the profile of the pulses.  The conclusion yielded
a "--measurement of the precession time consistent with the
predictions of general relativity--" \cite{stairs} and the authors
note that "--although the precision is, as yet, limited--", future
plans should allow for improvement.

\subsection{The double binary system PSR J0737-3039
\cite{lyne,burgay03}}

This is a highly significant discovery, due to the fact that it is
the only binary system discovered in which both neutron stars are
radio pulses; in particular it widens the scope for tests of
general relativity
\cite{kramer06} and for the study of the magnetospheres surrounding
pulsars.  This binary has an orbital period of 2.4 hours and the
periastion precession was observed to be
$16^{o}.9$/yr.  Pulsar A has a period of $22.7 ms$ whereas pulsar B
has a period of
$2.77s$, that is $A$ spins 122 times faster.  A detailed  listing
of measured and derived parameters are listed in
\cite{lyne,burgay03,kramer06}; in particular, in units of a solar
mass, $M=2.58708$ with estimated values of $m_{A}=1.3381$ and
$m_{B}=1.2489$.  This leads to
$\alpha_{g}\approx 4.4\times 10^{-6}$.  In addition, the
corresponding reduced mass $\mu$ is 0.64596.  Thus, using the
two-body spin precession formula
\cite{barker75,barker1975}, given in (\ref{gm217}) and
(\ref{gm218}), the excepted spin-orbit spin precession rates are
$4^{o}.8$/yr for pulsar A and $5^{o}.1$/yr for pulsar B (which
corresponds to geodetic precession periods of 75 years for A and 71
years for B).  Since the corresponding precession rate for the GP-B
earth is $1.838\times 10^{-3}$ degrees per year, we see that the
number for pulsar $A$ is larger by a factor of $2.7\times 10^{3}$. 
Despite these large values, they have not yet been observationally
verified.  However, it may be possible to deduce the effect of the
two-body spin-orbit term from the fact that it also results in a
value of
$\approx 4.06$ arc-sec/yr for the rate of precession of both the
angular momentum and the Runge-Lenz vector of the orbit about the
pulsar spin direction
\cite{oconnell04}.  This is reflected in a precession of the
inclination of the orbit \cite{oconnell04} and there is also a
contribution to the perihelion precession.  However, in the former
case, such an observation is not helped by the fact that the
orbital plane is close to the line of sight, the angle of
inclination being
$88^{o}.69$ \cite{kramer06}.  The latter case is thus probably more
promising since the perihelion precession has been measured to a
relative precision
$\approx 10^{-5}$.  We expect that the second-order
spin-independent post-Newtonian contribution
 to the periastron precession to be smaller than the lowest-order
Einstein-Robertson contribution by a factor
$\approx\alpha_{g}=4\times 10^{-6}$.  On the other hand, the
lowest-order spin-orbit contribution of the fast pulsar A will be
smaller by a factor, $F$ say, where

\begin{equation}
F=\frac{\omega^{(1)}}{\bar{\omega}}~\left(\frac{R}{a}\right)^{2},
\label{gm41}
\end{equation} where $\omega^{(1)}=44.0545 s^{-1}$ is the
precession frequency of pulsar A \cite{kramer06} and where
$R\approx 15kms$ is the radius of the pulsar A neutron star.  Thus,
since
$(\omega^{(1)}/\bar{\omega})=3.89\times 10^{3}$, and
$(R/a)=1.71\times 10^{-5}$, we see that
$F=1.1\times 10^{-4}$.  Thus, the lowest-order spin-orbit
contribution could be $\approx$ 10 times the second-order
spin-independent contribution and also the relative precision
$\approx 10^{-5}$.  This translates to a contribution to the
periastron precession $\approx \left(16.9\times
10^{-4}\right)^{o}/yr\approx 5$ arc-sec/yr, keeping in mind the
uncertainties underlying this calculation.  However, if the
observations provide even better relative precision in the future,
they could "--provide a way to obtain accurate information of the
moment of inertia of neutron stars--."
\cite{oconnell04}.

\section{Conclusions}

On the theoretical side, the general relativistic Hamiltonian for a
two-body system with spin, to order $c^{-2}$, contains spin-orbit
and spin-spin contributions, analogous to the corresponding
Hamiltonian for QED.  These terms separately manifest themselves
as spin and orbital precessions, which are in fact intimately
related, reflecting their common origin.  Also, it was convenient
from an observational point of view, to separate the orbital
precession into periastron, nodal and inclination angle
precessions.  Thus, a measurement of any of these precessions is
potentially a test of gravitomagnetism (spin-orbit and spin-spin
effects).

As we discussed above, there is already strong evidence (subsects.
3.2, 3.3 and 4.2) for the predicted spin-orbit effects in general
relativity.  This bears out the conclusion of a NASA 2003
scientific review who "--found "some erosion" in the scientific
value of the frame-dragging experiment--"
\cite{lawler03}.  However, it does appear, from the initial
announcement of specific results from the GP-B experiment that an
accuracy better than 1\% may have been achieved for the geodetic
effect but the lunar lasing experiment is already much better,
achieving an accuracy of 0.1\%.  Also, the other experiments, as
distinct from GP-B, have the advantage of being on-going
experiments.  For readers interested in other theories of
gravitation (for which I believe there is neither strong
theoretical, experimental or observational evidence), a detailed
discussion is given in
\cite{barker1976}.  In particular, it is clear that in all these
other theories, the intimate relationship between the spin-orbit
measurements of the GPB group and those of Ciufolini and Pavlis
\cite{ciu} still holds.

We emphasize that our considerations were confined to order
$c^{-2}$, which is an excellent approximation for the systems
considered for which the largest value of $\alpha_{g}$ is
$4.4\times 10^{-6}$.  However, motivated by the search for
gravitational radiation, there is widespread activity in
investigating the final evolution of a binary spinning black-hole
system; the results given in
\cite{barker75} provide a touchstone for sufficiently large
separations but strong field effects soon play a dominant role as
the merger takes place so that massive computational effects are
necessary to extract meaningful results \cite{campanelli07}.

Finally, we note that the theoretical work underpinning the results
presented here are treated in depth in a companion paper
\cite{oconnell}.  In particular, we discuss relevant conceptual
matters dealing with the relation between velocity and momentum
(especially for particles with spin), the non-uniqueness of spin
supplementary conditions and the choice of coordinates even at the
classical special relativistic level, the fact that a spinning
particle necessarily has a minimum radius, the corresponding
concepts in quantum theory (relating to such topics as the
Foldy-Wouthuysen transformation and position operators) and the
fact that the spin effects in quantum electrodynamics (obtained
from one-photon exchange) have their analogy in general relativity
(obtained from one-graviton exchange or purely classical
calculations) to the extent that except for (important) numerical
factors, the latter results may be obtained from the former by
simply letting
$e^{2}\rightarrow Gm_{1}m_{2}$.

\section{Post-Script}

A recent Comment by Kopeiken \cite{kope07} argues that " - - lunar
laser ranging (LLR) is not currently capable to detect
gravitomagnetic effects - - - ", based on the fact that a
Newtonian-like translation from the solar system barycentric (SSB)
frame to the geocentric frame (that is the frame co-moving with
the earth) gives rise to additional gravitomagetic terms which " -
- makes LLR insensitive to the gravitomagnetic interaction".  As
an example of such extra contributions, we note that coordinate
tranformations give rise to Thomas precession contributions of the
same order of magnitude as those arising from post-Newtonian
effects \cite{chan,ciufolini86}.  However, for the 3-body
earth-moon-sun system, there are several velocity-dependent
contributions and a Reply by Murphy by et. \cite{murphy2007} to
the Kopeiken Comment disputes his conclusion on the basis that " -
- - transformations of the velocity-dependent terms from one frame
to another - - - [are] strongly constrained by experiment".

More recently, Ciufolini in a lucid analysis \cite{ciufolini07},
reiterates a point which he made earlier in his book with Wheeler
\cite{ciufolini86} that " - - - the de Sitter effect and the
Lense-Thirring drag are fundamentally different phenomena - - - "
\cite{ciufolini86} since the curvature invariant $\ast$R.R is zero
in the former case (which is generated by the translational motion
of a mass) but non-zero in the latter case (which is generated by
the rotation of a mass).  Ciufolini then goes on to stress that "
- - - one cannot derive rotational gravitomagnetic effects from
translational ones, unless making additional theoretical
hypotheses such as the linear superposition of the gravitomagnetic
effects."  While we agree with Ciufolini, from our point of view,
both of the effects under discussion arise from a common spin-orbit
term in the basic Hamiltonian H, which can manifest itself in
various ways.  In particular, as detailed above, the same
spin-orbit term gives rise to both a precession of the spin (or
rotation axis) per se and also a precession of the perihelion of
the orbit.  In particular, the rate of precession of the spin
arising from the spin-orbit term in H depends on the angular
momentum and not on the spin whereas the rate of orbital
precession depends on the spin but not on the angular momentum. 
The rate of precession of the GP-B gyroscope orbiting the earth is
representative of the former case and so is the earth-moon
gyroscope orbiting the sun.  Concomitantly, in both cases, there
is an associated orbital precession; this is negligibly small in
the case of the GP-B gyro but not so for the case of the
earth-moon gyroscope because the latter is itself a two-body
system forming part of a three-body system.  As a consequence, the
motions of the earth and the moon can be separately tracked
leading, in particular, to spin-orbit amplitude contributions to
the lunar orbit at the 6 m level \cite {murphy07}.

We should also emphasize that, as already discussed, there are two
separate spin-orbit contributions.  However, in all one-body
problems, the dominant spin-orbit contribution to the spin
precession is associated with the spin of the smaller mass [see
(2.17) and (2.18)] i.e. the gyro mass in the case of GP-B and the
earth-moon mass in the case of LLR.  Of course, there is also a
spin-spin contribution in both cases; an attempt has been made to
measure this in the case of GP-B but it is generally
negligible for contributions to one-body orbital precessions
(compared to spin-orbit contributions and the dominant
Einstein-Robertson periastron precession term).  On the other
hand, for two-body systems, such as the double binary system, both
spin-orbit contributions to spin precession are important and, as
already noted, in Section 4.3, their contribution to the periastron
precession of the orbit of the double binary system may also be
measurable in further observations.

More generally, we emphasize that our analyses and calculations
demonstrate the universality of gravitational spin effects
regardless of whether the source is the spin of an elementary
particle, the rotation of a macroscopic body or a translational
effect.  I thank Professor Herbert Pfister for comments which
initiated this post-script. As a further update, we note that our
result for spin-orbit precession in a 2-body system \cite{barker75} has been verified,
to an accuracy of 13\%, by Breton et al., who observed a precession of the spin axis of pulsar B in the double binary system by an amount $4.77^{\circ}$/yr. \cite{breton08}.


\begin{thebibliography}{0}


\bibitem{oconnell74}  O'CONNELL R. F., \textit{Spin, Rotation and
C, P, and T Effects in the Gravitational Interaction and Related
Experiments, in Experimental Gravitation: Proceedings of Course 56
of the International School of Physics "Enrico Fermi"}, Varenna,
Italy, 1972, ed. B. Bertotti (Academic Press, 1974), pp. 496-514.

\bibitem{barker76}  BARKER B. M. and O'CONNELL R. F.,
\textit{General Relativistic Effects in Binary Systems, in Physics
and Astrophysics of Neutron Stars and Black Holes: Proceedings of
Course 65 of the International School of Physics "Enrico Fermi"},
Varenna, Italy, 1975, ed. R. Giacconi (North-Holland, 1976), pp.
437-447.

\bibitem{barker70}  BARKER B. M. and O'CONNELL R. F.,
\textit{Derivation of the Equations of Motion of a Gyroscope from
the Quantum Theory of Gravitation}, Phys. Rev. D \textbf{2} (1970)
1428.

\bibitem{schiff}  SCHIFF L. I.: Proc. Nat. Acad. Sci. \textbf{46}
(1960) 871; \textit{Proceedings of the Theory of Gravitation},
Jablonna, Poland, 1962, edited by L. INFELD (Paris, 1964), p. 71.

\bibitem{everitt}  EVERITT C. W. F., \textit{The Gyroscope
Experiment I: General Description and Analysis of Gyroscope
Performance, in Experimental Gravitation: Proceedings of Course 56
of the International School of Physics "Enrico Fermi"}, Varenna,
Italy, 1972, ed. B. Bertotti (Academic Press, 1974), pp. 331-360.  

\bibitem{brum}   BRUMFIEL G., Gravity probe falters, Nature
\textbf{444} (2006) 978.

\bibitem{hulse}  HULSE R. A. AND TAYLOR J. H., \textit{Discovery of
a Pulsar in a Binary System}, Astrophys. Journ. Lett.,
\textbf{195} (1975) L51.

\bibitem{barker75}   BARKER B. M. and O'CONNELL R. F., \textit{The
Gravitational Two Body Problem With Arbitrary Masses, Spins, and
Quadrupole Moments}, Phys. Rev. D \textbf{12} (1975) 329.  

\bibitem{barker1975}  BARKER B. M. and O'CONNELL R. F.,
\textit{Relativistic Effects in the Binary Pulsar PSR 1913+16},
Astrophys. J. Lett.
\textbf{199} (1975) L25.  

\bibitem{ciu} CIUFOLINI I. and PAVLIS, E.C., \textit{A
confirmation of the general relativistic prediction of the
Lense-Thirring effect}, Nature
\textbf{431} (2004) 958.

\bibitem{sitter76} DE SITTER W., \textit{On Einstein's Theory of
Gravitation and its Astronomical Consequences}, First Paper, Mon.
Not. R. Soc.
\textbf{76} (9) (1916) 699-728.

\bibitem{sitter72} DE SITTER W., \textit{On Einstein's Theory of
Gravitation and its Astronomical Consequences}, Second Paper, Mon.
Not. R. Soc.\textbf{ 76} (2) (1916) 155-184.

\bibitem{lense1}  LENSE J. and THIRRING H., \textit{$\ddot{U}$ber
den Einfluss der Eigenrotation der Zentralk$\ddot(o)$rper auf die
Bewegung der Planeten und Monde nach der Einsteinschen
Gravitationstheorie}, Phys. Z. 19 (1918)  pp. 156-163.

\bibitem{lense2} LENSE J. and THIRRING H., \textit{On the
gravitational effects of rotating masses:  The Thirring-Lense
papers}, trans. B. Mashhoon, F. W. Hehl, and D. S. Theiss, Gen.
Relativ. Gravit. \textbf{16} (1984) pp. 711-750.

\bibitem{will3} WILLIAMS J. G., TURYSHEV S. G., and BOGGS D. H.,
\textit{Progress in Lunar Laser Ranging Tests of Relativistic
Gravity}, Phys. Rev. Lett.
\textbf{93} (2004) 261101.

\bibitem{murphy07} MURPHY, JR. T.W. , NORDTVEDT K. AND TURYSHEV
S.G.,
\textit{Gravitomagnetic Influence on Gyroscopes and on the Lunar
Orbit}, Phys. Rev. Lett. \textbf{98} (2007) 071102.

\bibitem{stairs} STAIRS I. H., THORSETT S. E. and ARZOUMANIAN Z.,
\textit{Measurement of Gravitational Spin-Orbit Coupling in a
Binary-Pulsar System}, Phys. Rev. Lett.
\textbf{93} (2004) 141101.

\bibitem{lyne} LYNE A.G. et al., \textit{A Double-Pulsar System: A
Rare Laboratory for Relativistic Gravity and Plasma Physics},
Science 
\textbf{303} (2004) 1153.

\bibitem{oconnell05}  O'CONNELL R. F., \textit{A note on frame
dragging}, Class. Quantum Grav. \textbf{22} (2005) 3815.

\bibitem{barker84} BARKER B. M. and O'CONNELL R. F., \textit{The
Gyroscope Experiment}, Nature \textbf{312} (1984) 314.

\bibitem{oconnell85} O'CONNELL R. F., \textit{The Gyroscope Test
of Relativity}, Physics Today \textbf{38}, No.2 (1985) 104.

\bibitem{kahn07} KAHN B., Stanford press release, April 14, 2007.

\bibitem{stanford06} STANFORD, \textit{GP-B Mission Status
Updates}, Dec. 22, 2006, Sept. 2007 and Dec.2007.

\bibitem{reichhardt03} REICHHARDT T., \textit{Unstoppable Force},
Nature \textbf{426} (2003) 380.

\bibitem{wynne06} GWYNNE P., \textit{Make or break for gravity
experiment}, Phys. World (June 2006) 6.

\bibitem{collins04} COLLINS H., \textit{Gravity's shadow: the
search for gravitational waves}, (U. of Chicago Press) 2004.

\bibitem{bertotti87} BERTOTTI B., CIUFOLINI I., BENDER P. L.,
\textit{New Test of General Relativity: Measurement of de Sitter
Geodetic Precession Rate for Lunar Perigee}, Phys. Rev. Lett.
\textbf{58} (1987) 1062.

\bibitem{ciufolini86} CIUFOLINI I. and WHEELER J. A.,
\textit{Gravitation and Inertia}, (Princeton U. Press) 1986.

\bibitem{ciufolini04} CIUFOLINI I., PAVLIS E. C., and PERON R.,
\textit{Determination of frame-dragging using Earth gravity models
from CHAMP and GRACE}, New Astronomy Rev.
\textbf{11} (2006) 527.

\bibitem{ciufolini07} CIUFOLINI I.,
\textit{Dragging of Inertial Frames}, Nature \textbf{449} (2007)
41.

\bibitem{robertson38} ROBERTSON H. P., Ann. Math. \textbf{39}
(1938) 101.

\bibitem{kramer98} KRAMER M., \textit{Determination of the Geometry
of the PSR B1913+16 System by Geodetic Precession}, Ap. J.
\textbf{509} (1998) 856.

\bibitem{weisberg02} WEISBERG J. M. and TAYLOR J. H.,
\textit{General Relativistic Geodetic Spin Precession in Binary
Pulsar B1913+16: Mapping the Emission Beam in Two Dimensions}, Ap.
J. \textbf{576} (2002) 942.

\bibitem{burgay03} BURGAY M. et al., \textit{An Increased Estimate
of the Merger Rate of Double Neutron Stars from Observations of a
Highly Relativistic System}, Nature
\textbf{426} (2003) 531.

\bibitem{kramer06} KRAMER M. et al., \textit{Tests of General
Relativity from Timing the Double Pulsar}, Science \textbf{314}
(2006) 97.

\bibitem{oconnell04} O'CONNELL R. F., \textit{Proposed New Test of
Spin Effects in General Relativity}, Phys. Rev. Lett. \textbf{93}
(2004) 081103.

\bibitem{lawler03} LAWLER A., \textit{NASA Orders Make-or-Break
Tests for Gravity Probe}, Science \textbf{300} (2003) 880.

\bibitem{barker1976} BARKER B. M. and 	O'CONNELL R. F.,
\textit{Lagrangian-Hamiltonian formalism for the gravitational
two-body problem with spin and parametrized post-Newtonian
parameters $\gamma$ and $\beta$}, Phys. Rev. D \textbf{14} (1976)
861.

\bibitem{campanelli07} CAMPANELLI M., LOUSTO C. O., ZLOCHOWER Y.,
KRISHNAN B., and MERRITT D., \textit{Spin Flips and Precession in
Black-Hole-Binary Mergers}, Phys. Rev. D \textbf{75} (2007) 064030.

\bibitem{oconnell} O'CONNELL R.F., \textit{Spin and Rotation in
Physics}, in "Frame-dragging, gravitational-waves and
gravitational tests", ed. I. Ciufolini and R. Matzner, in honor of
J.A. Wheeler, to be published.

\bibitem{kope07} KOPEIKIN S.M., \textit{Comment on
"Gravitomagnetic Influence on Gyroscopes and on the Lunar Orbit"},
Phys. Rev. Lett. \textbf{98} (2007) 229001.

\bibitem{chan} CHAN LAI-HIM and O'CONNELL R.F., \textit{Two-body
problems-A unified, classical, and simple treatment of spin-orbit
effects}, Phys. Rev. D \textbf{15} (1977) 3058.

\bibitem{murphy2007} MURPHY, JR. T.W. , NORDTVEDT K. AND TURYSHEV
S.G.,
\textit{Reply to Kopeikin [41]}, Phys. Rev. Lett. \textbf{98}
(2007) 229002.

\bibitem{ciufolini07} CIUFOLINI I.. \textit{Gravitomagnetism,
Frame-Dragging and Lunar Laser Ranging}, arXiv:0704.3338v2 [gr-qc]
10 May 2007.

\bibitem{breton08} BRETON, R. P., Kaspi, V. M., Kramer, M., McLaughlin, M. A., Lyutikov, M., Ransom, S. M., Stairs, I. H., Ferdman, R. D., Camilo, F., Possenti, A., \textit{Relativistic Spin Precession in the Double Pulsar}, Science \textbf{321}, (2008) 104. 

\end{thebibliography}
\end{document}